\documentstyle[preprint,eqsecnum,prd,aps,epsfig]{revtex} 
 
\begin{document} 
 
\draft 
 
\title{Timelike and null focusing singularities in spherical symmetry: a
solution to the cosmological horizon problem and a challenge to the cosmic 
censorship hypothesis}

\author{Marie-No\"elle C\'el\'erier} 
 
\address{LUTH, Observatoire de Paris-Meudon, 5 place Jules Janssen, 
92195 Meudon Cedex, France} 
 
\author{Peter Szekeres} 
 
\address{Department of Physics and Mathematical Physics, University of 
Adelaide, GPO Box 498, Adelaide SA 5001, Australia}

\date{\today} 
 
\maketitle

\begin{abstract} 
 
Extending the study of spherically symmetric metrics 
satisfying the dominant energy condition and exhibiting singularities 
of power-law type initiated in \cite{SI93}, we identify two classes 
of peculiar interest: focusing timelike singularity solutions with the 
stress-energy tensor of a radiative perfect fluid (equation of state: 
$p={1\over 3} \rho$) and a set of null singularity classes verifying identical 
properties. We consider two important applications of these results: 
to cosmology, as regards the possibility of solving the horizon 
problem with no need to resort to any inflationary scenario, and to 
the Strong Cosmic Censorship Hypothesis to which 
we propose a class of physically consistent counter-examples. 
 
\end{abstract}

 

\section{Introduction} 
\label{int}

In a recent work, devoted to some developments of the ``Delayed Big-Bang'' 
(DBB) cosmological model \cite{MNC00}, it was shown that 
any cosmological model exhibiting a null singularity surface is naturally 
and permanently free from any horizon problem. This property can easily be 
extended to models with a timelike singularity, as we shall see in 
Sect.\ref{sol}.\\

This important issue of the horizon problem in cosmology has induced us 
to investigate the constraints which can be imposed on 
the stress-energy tensor by the requirement of a non-spacelike singularity. 
Even though this requirement is only a sufficient, but not 
necessary, condition for the resolution of the problem (see discussion 
in Sect.\ref{sol}), it is an interesting cosmological question in its own 
right. \\

Part of the work has already been done by Szekeres and Iyer \cite{SI93} 
(hereinafter referred to as SI), who investigated, in the spherically 
symmetric case, the constraints imposed by the requirement of a timelike 
singularity, with the object of exploring the validity of the Strong Cosmic 
Censorship Hypothesis (SCCH). Albeit the situation considered in SI is a 
collapse, the results can be extended straightforwardly to the cosmological 
case by reversing the direction of time. \\

In this paper, we still concentrate our attention on spherically 
symmetric singularities. Despite this specialization, 
interesting information can undoubtedly be gained, and any results obtained 
should be usefully applied to both the cosmological issue and the SCCH. \\ 
 
Until now, our preliminary works aiming at solving the horizon 
problem without recourse to the inflationary paradigm 
\cite{MNC00,CS98,SC99} retain the simplifying approximation of a 
dust dominated universe, the DBB model. However, as stressed in 
\cite{CS98}, when going backward on the light-cones issuing from the 
last-scattering surface towards the singularity, the energy density 
increases and one would expect on physical grounds that the 
pressure should do likewise. Thus, the radiation becomes the 
dominant component in the universe, and possibly a relativistic 
equation of state such as $P={1\over 3}\rho$ should apply. In the 
present article we show that such an equation of state is 
compatible with a non-spacelike singularity, providing therefore 
new physically consistent models for a horizon problem free 
primordial universe. \\

Furthermore, the consideration of a special case which, for no 
good reason, has been omitted from the analysis given in SI yields 
an example of focusing timelike singularity compatible with the above 
radiative equation of state and at variance with the SCCH. \\

In Sec.~\ref{sol} we review the way a non-spacelike singularity leads to the 
resolution of  the standard cosmological horizon problem. In Secs.~\ref{cits} 
to ~\ref{ns}, we identify the constraints imposed on the stress-energy tensor 
by the requirements of a timelike and a null singularity. Secs.~\ref{cissdm} 
and ~\ref{scch} are devoted to the application of the obtained results to 
the cosmological horizon problem and the SCCH issue, respectively. The 
conclusions are stated in Sec.~\ref{concl}. A derivation of the timelike 
character of spherically symmetric shell-crossing surfaces is proposed 
in the Appendix.

\section{Solving the horizon problem} 
\label{sol}

As shown in \cite{MNC00}, the horizon problem develops sooner or 
later in any cosmological model exhibiting a spacelike singularity such as 
that occurring in standard FLRW universes. 
Simply stated, the horizon problem is this: In hot Big-Bang models the 
comoving region over which the cosmic microwave background radiation (CMBR) 
is observed to 
be homogeneous to better than one part in $10^5$ at the last-scattering 
surface 
is much larger than the intersection of this surface with future light-cone 
from the ``Big-Bang''. 
As this light-cone provides the maximal distance over which causal processes 
could have 
propagated since a given point on the ``Big-Bang'', the observed isotropy 
of the CMBR remains unexplained. \\

\begin{figure} 
\centering 
\includegraphics[height=4cm,width=6.87cm,angle=-90]{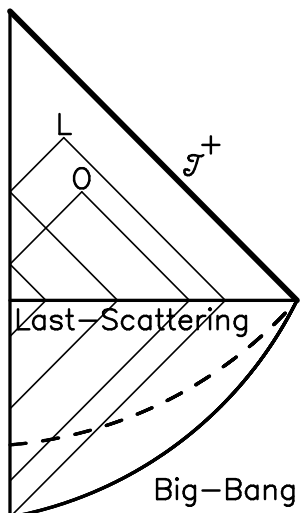} 
\caption{Penrose-Carter diagram showing the horizon problem 
in a universe with spacelike singularity.} 
\label{f:1} 
\end{figure}

Even inflation only postpones the occurrence of the horizon 
problem since it does not change the spacelike character of the 
singularity and is insufficient to solve it permanently. 
This is shown in Fig.~\ref{f:1}, where thin lines represent light-cones and 
the CMBR as seen by an observer $O$ corresponds to the 
intersection of the observer's backward light-cone with the 
last-scattering line. For a complete causal connection to occur between 
every pair of points in this intersection segment, backward light signals 
issuing from points therein must reach the vertical axis before they reach 
the spacelike ``Big-Bang'' curve. $L$ is thus a limiting event 
beyond which any observer experiences the 
horizon problem. Adding an inflationary phase in the primordial history of 
the universe amounts to adding a slice of de Sitter space-time, indicated 
here by the region between the dashed line and the ``Big-Bang''. The effect 
of this region is 
merely to postpone the event $L$, allowing the current observer $O$ to see a 
causally connected 
CMBR. At later times the observer reaches the region above $L$ and the 
horizon problem reappears.

In \cite{MNC00} a permanent solution to this problem was proposed, using the 
DBB class of models, valid for all 
observers regardless of their location in the universe. 
These models have a non-spacelike singularity which can arise, for example, 
as shell-crossings 
(see \cite{SL99} for a detailed characterization of a shell-crossing 
singularity). In \cite{MNC00} 
shell-crossings were mistakenly claimed to be null surfaces, whereas they 
are in fact timelike. 
A derivation of this property, valid for general spherically symmetric 
models, is given in the Appendix. 
Fig.~\ref{f:2} shows that a non-spacelike singularity always gives rise to 
an everywhere 
causally connected model of universe. Every pair of points in the CMBR seen 
by the current observer $O$ 
are causally connected since a past light signal from any point in the 
segment of 
the last-scattering surface seen by $O$ reaches the vertical axis before 
arriving at the 
null (straight line) or timelike (curved line) singularity. 
The same holds for any event $O'$ in the observer's past or future. \\ 
 
Therefore, any cosmological model exhibiting 
the equation of state of radiation near a non-spacelike (i.e., 
timelike or null) singularity, whatever its 
type (shell-cross or focus), could be considered as a 
physically consistent candidate to represent the primordial universe. 
 
\begin{figure} 
\centering 
\includegraphics[height=4.07cm,width=8.76cm,angle=-90]{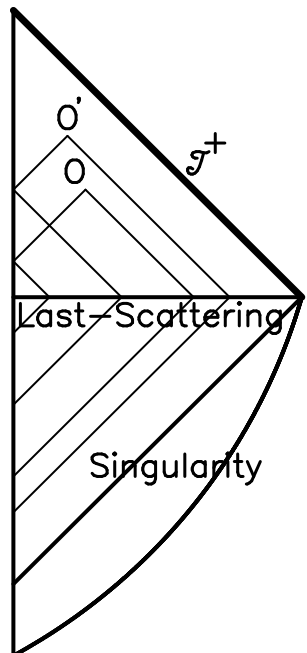} 
\caption{Penrose-Carter diagram showing causal connectedness of a 
universe with non-spacelike singularity.}
\label{f:2}  
\end{figure} 
 
\section{Double null coordinates and singularities of power-law type} 
\label{cits}

In SI spherically symmetric metrics were studied, having the form 
\begin{equation} 
ds^2=-dt^2+[t-\tau(r)]^{2a}f^2(r,t)dr^2+[t-\tau(r)]^{2b}g^2(r,t)d\Omega^2, 
\label{msp} 
\end{equation} 
where $f$ and $g$ are functions of $r$ and $t$ which are regular and 
non-vanishing at the singularity surface $t=\tau(r)$. 
This type of singularity was said to be ``of power-law type'' and is present 
in spherically symmetric dust solutions 
(LTB metrics \cite{GL33,RT34,HB47}), 
all FLRW perfect fluid solutions, and cosmologies with singularities of the 
Lifshitz-Kalatnikov type \cite{LK63}. 
We will be proposing a slightly different and more general definition of this 
concept. 
While the only purpose of SI was to discuss the validity of the cosmic 
censorship 
hypothesis by considering the behavior of collapsing space-times near 
the singularity, most of the results of that paper can readily be generalized 
to a cosmological setting by reversing the direction of time. 
As the present paper is mostly directed at cosmological issues, we are mainly 
interested in the region $t>\tau(r)$, and have made an appropriate sign change 
in Eq.~(\ref{msp}) to that used in SI. 
The newly obtained results will again be applied to 
the SCCH issue, by reversing the direction of time. \\

The method adopted in SI is to pass to double null coordinates 
$u(r,t)$ and $v(r,t)$ such that 
\begin{equation} 
ds^2=-2e^U dudv+e^Vd\Omega^2. 
\label{mdn} 
\end{equation} 
Such coordinates have a certain rigidity, in that the only 
available coordinate freedoms are of the form 
\begin{equation} 
u'=\mu(u,)\qquad v'=\nu(v), 
\label{cof} 
\end{equation} 
where $\mu$ and $\nu$ are arbitrary differentiable functions. 
Radial null lines ($\theta=\text{const}$, $\phi=\text{const}$) are exactly the 
curves $u=\text{const}$ or $v=\text{const}$. We assume the future pointing 
directions on 
these null lines are those with increasing values of $v$ and $u$ respectively. 
 
In SI a coordinate transformation converting the metric 
of Eq.~(\ref{msp}) into the double null coordinates of Eq.~(\ref{mdn}) 
is performed in a neighborhood of $r=r_0,\; 
t=\tau(r_0)$ 
by carrying out a series expansion of the form 
\begin{eqnarray} r &=& r_0 +u + f_1(u) x^{a_1} +f_2(u) x^{a_2} +\dots 
\label{r-uv} \\ 
t &=& \tau(r_0 +u) + g_1(u) x^{b_1} +g_2(u) x^{b_2} +\dots \label{t-uv}\\ 
&=& \tau(r_0)-\tau'(r_0+u)f_1(u)x^{a_1} + g_1(u)x^{b_1} + \dots \label{t-uv1} 
\end{eqnarray} 
where $ 0<a_1 <a_2 <\dots$, and $0<b_1 <b_2 <\dots$. 
By Eq.~(\ref{t-uv1}) the singularity at $t=\tau(r)$ occurs at $x=0$, and the 
freedom of Eqs.~(\ref{cof}) can be used to 
express the function $x(u,v)$ in the form 
\begin{equation} \label{xuv} x=lu+k v \qquad \text{where } l,k = \pm 1 
\text{ or } 0.\end{equation} 
The signs of $l$ and $k$ should be chosen such that $x>0$ for $t>\tau(r)$ in 
the neighborhood of the singularity. 
It is easy to verify that the singularity $x=0$ has spacelike character if 
$lk = 1$, timelike if $lk = -1$, 
and null if $lk = 0$. 
For example, if $lk=1$, then in the limit as $x\rightarrow 0$, the surface $x=\text{const}>0$ 
intersects both null lines 
$u=\text{const}$ and $v=\text{const}$ in positive (future values) if $l=k=1$, 
while it intersects them in the past if $l=k=-1$. 
The first case $x=u+v$ therefore corresponds to a spacelike singularity in 
the past (cosmological), while $x=-u-v$ represents a spacelike singularity 
in the future (collapse). A similar analysis for the case $lk=-1$ results 
in one ingoing null ray and the other outgoing. It therefore is timelike, 
which can be though of as having both a cosmological and collapse character. 
When $lk=0$ one of the rays $u=0$ or $v=0$ is tangential to the singularity 
surface, which can be thought of as null. \\ 
 
After some analysis, the functions $U$ and $V$ appearing in Eq.~(\ref{mdn}) 
can be shown to have the form 
\begin{equation} e^U = 2 t_u t_v = x^p e^{\alpha},\qquad e^V = (t-\tau)^{2b} 
g^2 = x^q e^{\beta} \label{eUV}\end{equation} 
where 
\begin{equation}\label{alpha} \alpha = \alpha_0(u) + \alpha_1(u)x^{p_1} 
+ \dots, \end{equation} 
\begin{equation}\label{beta} \beta = \beta_0(u) + \beta_1(u)x^{q_1} + \dots , 
\end{equation} 
and the exponents $p$ and $q$ depend on $a$ and $b$ in a variety of ways 
detailed in SI. Exponents $p_1,\,p_2, \dots,q_1,, \dots$ appearing in the 
expansions of $\alpha$ and $\beta$ can also be evaluated in principle from 
the exponents $a_1, a_2$,... $b_1, b_2$,... occuring in Eqs.~(\ref{r-uv}) 
to (\ref{t-uv1}), though it may be difficult to give general expressions 
for them. \\ 
 
The following two examples should give an idea of the kind of results expected: 
\begin{trivlist} 
\item{\bf Einstein-de Sitter dust solution:}\\[1ex] 
These are only spatially flat dust solutions of the form given in 
Eq.~(\ref{msp}) having 
$\tau(r)=\text{const}.=0$. The metric can be written 
\begin{equation} 
ds^2 = -dt^2+t^{4\over3}(dr^2+r^2d\Omega^2) 
\label{mes} 
\end{equation} 
and the singularity is well-known to be spacelike (since future pointing null 
geodesics emanate from $t=0$  in all directions). It is for this reason that 
the horizon problem occurs in cosmological models of this type. \\ 
 
The singularity has $a=b={2\over 3}$, and the transformation to double null 
coordinates is straightforward to perform. The series expansion has $a_1=1$ 
and $b_1=3$ and gives rise to the following 
exponents in Eqs.~(\ref{eUV}), ~(\ref{alpha}) and ~(\ref{beta}): 
\[p=q=4,\qquad p_1=q_1=1, \quad \dots \] 
 
\item{\bf Zero energy LTB solutions:} 
\begin{equation} 
ds^2 = -dt^2+(t-t_0(r))^{-{2\over3}}(t-t_1(r))^2dr^2 
+r^2(t-t_0(r))^{4\over3}d\Omega^2, 
\label{mtb} 
\end{equation} 
where $t=t_0(r)$ is the focusing singularity and $t=t_1(r)= 
t_0(r)+{2\over3}rt'_0(r)$ is a shell-crossing singularity. \\ 
 
A $t_0$ singularity (with $t'_0(r_0)\neq 0$) has $a=-{1\over 3}$, 
$b={2\over 3}$, and the exponents in the power series of Eqs.~(\ref{r-uv}) 
and ~(\ref{t-uv}) turn out to be \[a_1=1,\quad a_2={5\over 4}, \quad \dots 
\qquad b_1={3\over 4}, \quad b_2=1, \dots \] 
and give rise to the following exponents in the expansion of $U$ and $V$: 
\[p=\frac{2a}{1-a}=-{1\over 2}, \quad q= \frac{2b}{1-a} = 1, 
\qquad p_1=q_1={1\over 2}, \dots\] 
In this case it turns out that one must have $l=k=1$ as in the 
Einstein-de Sitter case and the singularity is spacelike. \\ 
 
A $t_1$ singularity (with $t'_1(r_0)\neq 0$) has $a=1$ and $b=0$, 
with transformation exponents \[a_1={1\over2},\quad \dots  \quad b_1=1, 
\quad \dots, \] 
and 
\[p=q=0,\qquad p_1=q_1=1, \qquad p_2=q_2={3\over 2}, \quad \dots \] 
The singularity in this case has $lk=-1$, is timelike (see the Appendix) 
and have no horizon problem. Accordingly, the dust DBB models have been 
constructed in such a way that the first singularity encountered when 
going backward on timelike curves is of type $t_1$. 
 
\end{trivlist} 
 
Although the singularity surface $x=0$ of a metric of the type described 
by Eq.~(\ref{msp}) 
is generally non-null, there is no reason to impose this restriction when 
starting from the double null form of the metric of Eq.~(\ref{mdn}). 
We will therefore define the singularity surface of a spherically symmetric 
metric as being {\em of power-law type} if it can be expressed in the form 
of Eq.~(\ref{mdn}) with 
\begin{eqnarray} 
U &=& p\ln x+\alpha_0(u)+\alpha_1(u)x^{p_1}+\alpha_2(u)x^{p_2}+\dots, \label{ude}\\ 
V &=& q\ln x+\beta_0(u)+\beta_1(u)x^{q_1}+\beta_2(u)x^{q_2}+ \dots, \label{vde} 
\end{eqnarray} 
where $0<p_1<p_2...,\quad 0<q_1<q_2<...$ and $x$ is given by Eq.~(\ref{xuv}). 
As there are essentially no further coordinate freedoms, this definition 
is invariant. In case the reader wonders why the functions $\alpha_1$, 
$\beta_1$ are not postulated to be regular functions of both variables $u$ 
and $v$, it is simple to express $v$ as either $u\pm x$ or $-(u\pm x)$, 
substitute in the functions and expand as power series.  The result would 
then be that given in Eqs.~(\ref{ude}) and (\ref{vde}).

\section{Stress-energy tensor near timelike singularities of power-law type} 
\label{ts}

The Einstein tensor for the metric of Eq.~(\ref{mdn}) has the following 
non-vanishing 
components (setting $x^0=u$, $x^1=v$, $x^2=\theta$, $x^3 =\phi$): 
\begin{eqnarray} 
G^{0}_{0} &=&  G^{1}_{1} = -e^{-V} - e^{-U}\bigl(V_{01} +V_0 V_1\bigr) \nonumber \\ 
&=& -x^{-q} e^{-\beta} - e^{-\alpha}x^{-p}\biggl(\frac{kl(q^2 - q)}{x^2} 
+\frac{k\beta_u +l\beta_v}{x} + \beta_{uv} +\beta_u \beta_v \biggr) 
\label{Gtens1} \\ 
G^{0}_{1} &=& e^{-U}\bigl(V_{11} -U_1 V_1 + \frac{1}{2} V_{1}^{\;2}\bigr) \nonumber \\ 
&=& -e^{-\alpha}x^{-p}\biggl(\frac{k^2(q+pq-\frac{1}{2} q^2)}{x^2} 
+\frac{k((p-q)\beta_v +q\alpha_v)}{x} 
-\beta_{vv} + \alpha_v \beta_v - \frac{1}{2} \beta_{v}^{\;2}\biggr) 
\label{Gtens2} \\ 
G^{1}_{0} &=& e^{-U}\bigl(V_{00} -U_0 V_0 + \frac{1}{2} V_{0}^{\;2}\bigr) \nonumber \\ 
&=& -e^{-\alpha} x^{-p} \biggl(\frac{l^2(q+pq-\frac{1}{2} q^2)}{x^2} 
+\frac{l((p-q)\beta_u +q\alpha_u)}{x} 
-\beta_{uu}+\alpha_u \beta_u - \frac{1}{2} \beta_{u}^{\;2}\biggr) 
\label{Gtens3} \\ 
G^{2}_{2} &=& G^{3}_{3} = -e^{-U}\bigl(U_{01} + V_{01} 
+ \frac{1}{2} V_0 V_1 \bigr)  \nonumber \\ 
&=& -e^{-\alpha}x^{-p}\biggl( \frac{lk(\frac{1}{2} q^2 - p - q)}{x^2} 
+\frac{q(k\beta_u +l\beta_v)}{2x} 
+\alpha_{uv}+ \beta_{uv} + \frac{1}{2} \beta_u \beta_v \biggr) 
\label{Gtens4} 
\end{eqnarray} 
 
The stress-energy tensor arising from Einstein's equations 
\[T^{\mu}_{\;\nu}= G^{\mu}_{\;\nu}\] 
has the form 
\[T^{\mu}_{\;\nu} =\rho u^\mu u_\nu + P_r f^\mu f_\nu 
+ P_{\perp} h^{\mu}_{\;\nu}, \] 
where $u^\mu$ is the unit timelike eigenvector and $f^\mu$ the unit 
spacelike radial eigenvector, \[T^{\mu}_{\;\nu} u^\nu= -\rho u^\mu, 
\qquad T^{\mu}_{\;\nu} f^\nu = P_r f^\mu \] having components 
\[u^\mu= \Bigl(u^0,\: u^1 = u^0 \sqrt{\frac{G^{1}_{0}}{G^{0}_{1}}}, 
\: 0, \: 0 \Bigr), \qquad g_{\mu\nu} u^\mu u^\nu = -2e^U u^0u^1 = -1,\] 
\[f^\mu= \Bigl(f^0,\: f^1 = -f^0 \sqrt{\frac{G^{1}_{0}}{G^{0}_{1}}}, 
\: 0, \: 0 \Bigr), \qquad g_{\mu\nu} f^\mu f^\nu = -2e^U f^0 f^1 = 1,\] 
and $h^{\mu}_{\;\nu}$ is the projection tensor into the space orthogonal 
to $u^\mu$ and $f^\mu$, 
\[h^{\mu}_{\;\nu} = \delta^{\mu}_{\;\nu} + u^\mu u^\nu -f^\mu f^\nu, \] 
whose only non-vanishing components are $h^{2}_{\;2} = h^{3}_{\;3} = 1$. 
In order for the eigenvectors $u^\mu$ and $f^\mu$ to be real it is 
necessary that $G^{1}_{0}$ and $G^{0}_{1}$ have the same sign. 
The density and radial pressure are given by 
\[\rho=-G^0_0 -G^0_1 \sqrt{\frac{G^1_0}{G^0_1} }, \qquad 
P_r=G^0_0 -G^0_1 \sqrt{\frac{G^1_0}{G^0_1} }, \] 
while tangential pressure is given by 
\begin{equation}\label{Pt}P_{\perp} = G^2_2 = G^3_3. \end{equation} 
It is common to impose the dominant energy condition $\rho > |P|$ as 
a physical requirement on the system, where $P$ is the pressure in any 
direction (and discounting the ``extreme'' case $\rho = \pm P$). 
Using $\rho+P_r >0$ we obtain 
\begin{equation} \label{G01ineq} G^0_1 < 0, \qquad G^1_0< 0. \end{equation} 
and the density and radial pressure are given by 
\begin{equation} \label{rhoPr} \rho  = -G^0_0 + \sqrt{G^0_1 G^1_0}, \qquad 
P_r= G^0_0 + \sqrt{G^0_1 G^1_0}. \end{equation} 
The dominant energy conditions $\rho > |P_r|$, $\rho > |P_{\perp}|$ 
imply the further inequalities 
\begin{equation} \label{G00ineq} G^0_0 < 0 , \qquad  |G^2_2 |< -G^0_0 
+ \sqrt{G^0_1 G^1_0},  \end{equation} 
which also guarantee positive density, $\rho > 0$. \\ 
 
For a perfect fluid with a barotropic equation of state, we have 
$P=P_r=P_{\perp}$ with 
\[P = \gamma\rho,\qquad -1<\gamma<1 \] 
and substitution in Eqs.~(\ref{Pt}) and (\ref{rhoPr}) results in 
\begin{equation} G^2_2 = G^3_3 = \frac{2\gamma}{\gamma -1} G^0_0, 
\qquad \sqrt{G_1^0G_0^1}= \frac{\gamma+1}{\gamma-1} G^0_0 
\qquad (G^0_0 <0). \label{gii}  \end{equation} 
For radiation, $\rho = \frac{1}{3} P$, we have 
\[G^2_2 = -G^0_0, \qquad  \sqrt{G^0_1 G^1_0} = -2 G^0_0.\] 
 
From Eqs.~(\ref{Gtens1}-\ref{Gtens3}) and (\ref{rhoPr}) we see that 
if $\rho \rightarrow \infty$ as $x\rightarrow 0$, then $q>0$ or $p>-2$. 
If the pressure is non-extreme in this limit, $P_r \gtrsim 
-\rho$ as $x\rightarrow 0$ then $q\leq p+2$. 
The only way these conditions are consistent is if 
\[p>-2, \qquad q\leq p+2.\] 
A detailed discussion for the case $kl\neq 0$ (timelike or spacelike 
singularity) and $q<p+2$ results in the following conclusion of SI: 
 
\begin{trivlist} 
\item{\em A timelike singularity of power-law type, in whose neighborhood 
the energy-stress tensor satisfies the dominant energy condition, must 
either 
\begin{enumerate} 
\item be a (dustlike, $P=0$) shell-cross singularity, or 
\item have an asymptotically extreme equation of state ($|P_r|\approx \rho$ 
or $|P_{\perp}|\approx \rho$), or 
\item possess a negative pressure ($P_r<0$ or $P_{\perp}<0$) in its 
neighborhood. 
\end{enumerate}} 
\end{trivlist} 
 
However the case $q=p+2$ has, for no good reason, been omitted in the 
analysis given in SI. We now give details of this case.

\subsection*{The case $ {\bf kl= \pm 1, \quad q=p+2, \quad p>-2}$} 
 
Since we must have $q>0$ in this case, the dominant behaviour of the 
various components of $G^\mu_\nu$ as $x\rightarrow 0$ is, on setting 
$\varepsilon = kl =\pm 1$, 
\begin{eqnarray} 
G^0_0 &\approx& -x^{-q} \Bigl( e^{-\beta_0} + \varepsilon e^{-\alpha_0} 
(q^2-q) \Bigr) \label{speccase1} \\ 
G^1_0 &\approx& G^0_1 \approx -x^{-q} e^{-\alpha_0}\frac{1}{2} q(q-2) 
\label{speccase2}\\ 
G^2_2 &\approx& -x^{-q} e^{-\alpha_0}\varepsilon \frac{1}{2} (q-2)^2 
\label{speccase3} 
\end{eqnarray} 
By Eq. (\ref{G01ineq}) and $q>0$ it follows that $q>2$, and using 
Eqs. (\ref{Pt}) and (\ref{rhoPr}), we have 
\begin{eqnarray} \rho &\approx& x^{-q}\Bigl( e^{-\beta_0} + e^{-\alpha_0} 
\frac{q}{2}\bigl(2\varepsilon(q-1) +q-2 \bigr) \Bigr) \label{rhospec}\\ 
P_r &\approx& x^{-q}\Bigl( -e^{-\beta_0} + e^{-\alpha_0} \frac{q}{2} 
\bigl(-2\varepsilon (q-1) +q-2 \bigr) \Bigr) \label{Prspec}\\ 
P_{\perp} &\approx& -x^{-q} e^{-\alpha_0}\varepsilon \frac{(q-2)^2}{2}. 
\label{Ptspec} 
\end{eqnarray} 
 
For the case of a spacelike singularity $\varepsilon=1$, 
\[\rho \approx x^{-q} \Bigl( e^{-\beta_0} + e^{-\alpha_0} 
\frac{q(3q-4)}{2}\Bigr) \: > \: 0 \qquad \text{since } q>2.\] 
However pressures in both radial and tangential directions are 
negative, 
\begin{eqnarray*} 
P_r &\approx& x^{-q}\Bigl(-e^{-\beta_0} - e^{-\alpha_0} \frac{q^2}{2} 
\Bigr) \: < \: 0, \\ 
P_{\perp} &\approx& -x^{-q} e^{-\alpha_0} \frac{(q-2)^2}{2} \: < \: 0. 
\end{eqnarray*} 
Thus, while the dominant energy conditions 
\[\rho+P_r=x^{-q}e^{-\alpha_0}q(q-2) >0\] 
and 
\[\rho+P_{\perp}= x^{-q}\Bigl(e^{-\beta_0} +e^{\alpha_0} 
\bigl(q^2 -2\bigr) \Bigr)>0 \] 
clearly hold for $q>2$, the negative pressures do not allow for a 
radiation limit. \\ 
 
In the case of a timelike singularity, $\varepsilon = -1$, we have 
\begin{eqnarray*} 
\rho &\approx& x^{-q}\Bigl( e^{-\beta_0} - e^{-\alpha_0} 
\frac{q^2}{2}\Bigr), \\ 
P_r &\approx& x^{-q}\Bigl(-e^{-\beta_0} + e^{-\alpha_0} 
\frac{q(3q-4)}{2}\Bigr), \\ 
P_{\perp} &\approx& x^{-q} e^{-\alpha_0} \frac{(q-2)^2}{2}  > 0. 
\end{eqnarray*} 
The positive density condition, $\rho>0$, gives 
\[e^{-\beta_0}>e^{-\alpha_0} \frac{q^2}{2}\] 
from which, using $q>2$, it is possible to verify the radial dominant 
energy inequality $G^0_0 <0$. Setting 
\[e^{-\beta_0}=e^{-\alpha_0}(\frac{q^2}{2} + C_0(u)) \qquad 
\text{where }C_0(u)>0,\] 
we have 
\begin{eqnarray} 
\rho &\approx& x^{-q} e^{-\alpha_0} C_0, \label{rhospec2}\\ 
P_r &\approx& x^{-q} e^{-\alpha_0}\bigl(-C_0 + q^2-2q \bigr), 
\label{Prspec2} \\ 
P_{\perp} &\approx& x^{-q} e^{-\alpha_0} \frac{(q-2)^2}{2} . 
\label{Ptspec2} 
\end{eqnarray} 
Essentially any sensible equation of state can be obtained in the 
vicinity of the singularity $x=0$ by a judicious choice of the function 
$C_0(u)$. For example, in the case of a perfect fluid (isotropic pressure), 
\[P_r = P_{\perp} \;\Longrightarrow \; C_0(u) = \frac{q^2}{2}-2 >0\] 
and 
\[\gamma = \frac{P}{\rho} = \frac{q-2}{q+2} >0 .\] 
A radiative equation of state, $\gamma = \frac{1}{3}$ is achieved if 
\[q=4,\qquad C_0(u) = 6.\] 
 
\section{Null singularities} 
\label{ns} 
 
The case where $x=0$ is a null singularity, $kl=0$ is not considered in SI, 
and should not be discarded without further investigation. In this case the 
dominant energy condition with $\rho \not\approx -P_r$ as $x\rightarrow 0$, 
together with $\rho \rightarrow \infty$ implies that 
\begin{equation}\label{pqineq}q\leq p+1,\qquad  p>-1. \end{equation} 
There are three essential cases to consider. 
 
\subsection*{(i) ${\bf \qquad l=1, \quad k=0}$ } 
 
Since $x=u$ in this case the particular choice of series expansion in 
Eqs.~(\ref{ude}) and (\ref{vde}) means that both functions $U$ and $V$ 
appearing in the double null coordinate form of the metric are functions 
of $u$ alone, $U=U(u)$, $V=V(u)$. Hence $\alpha_v= \beta_v = \beta_{vv}= 0$ 
in Eqs.~(\ref{Gtens1})-(\ref{Gtens4}), and consequently $G^0_1 =0$. 
By using Eq.~(\ref{rhoPr}) we arrive at the physically unacceptable 
condition $\rho = -P_r$ in the neighborhood of the singularity. 
 
\subsection*{(ii) ${\bf  \qquad l=0, \quad k=1, \quad p>-1, \quad q<p+1.}$ } 
 
In this case $x=v$ and the principal terms in Eqs.~(\ref{Gtens1})- 
(\ref{Gtens4}) are 
\begin{eqnarray*} 
G^0_0 &\approx& -v^{-p-1} e^{-\alpha_0} \beta'_0(u), \\ 
G^0_1 &\approx& -v^{-p-2} e^{-\alpha_0}q(1+p-\mbox{$\frac{1}{2}$} q), \\ 
G^1_0 &\approx& -v^{-p} e^{-\alpha_0}( -\beta''_0 +\alpha'_0 \beta'_0 
-\mbox{$\frac{1}{2}$} (\beta'_0)^2) ,\\ 
G^2_2 &\approx& -v^{-p-1} e^{-\alpha_0} \frac{q}{2} \beta'_0. 
\end{eqnarray*} 
The inequality $G^0_1<0$ implies $q(1+p-\frac{1}{2}q) >0 $. 
Hence, if $q<0$ then $p<\frac{1}{2} q-1 <-1$, contradicting the stated 
condition $p>-1$. Thus $q>0$. On the other hand, the inequality $G^0_0<0$ 
gives $\beta'_0 >0$, and the tangential pressure must be negative, 
$P_{\perp}=G^2_2 <0$.  This certainly does not permit an isotropic radiative 
fluid to be present near a null singularity of this type.  \\ 
 
We are left to consider one final case: 
 
\subsection*{(iii) ${\bf \qquad l=0, \quad k=1, \quad p>-1, \quad q=p+1.}$ } 
 
The components of the Einstein tensor are asymptotically dominated by the 
following terms: 
\begin{eqnarray*} 
G^0_0 &\approx& -v^{-p-1}  \Bigl(e^{-\beta_0} + e^{-\alpha_0} \beta'_0(u) 
\Bigr), \\ 
G^0_1 &\approx& -v^{-p-2} e^{-\alpha_0}\frac{q^2}{2}, \\ 
G^1_0 &\approx& -v^{-p} e^{-\alpha_0}\bigl( -\beta''_0 +\alpha'_0 \beta'_0 
-\mbox{$\frac{1}{2}$} (\beta'_0)^2 \bigr) ,\\ 
G^2_2 &\approx& -v^{-p-1} e^{-\alpha_0} \frac{q}{2} \beta'_0. 
\end{eqnarray*} 
By the inequalities ~(\ref{G01ineq}) and ~(\ref{G00ineq}) we may set 
\[e^{-\beta_0} = - e^{-\alpha_0} \beta'_0(u) +e^{-\alpha_0} A_0(u) \qquad 
\text{where } A_0(u)>0,\] 
and 
\[-\beta''_0 + \alpha'_0\beta'_0 -\frac{1}{2} (\beta'_0)^2 = 2B^2_0(u) 
\qquad \text{where } B_0(u)>0.\] 
Density and pressure components are found from Eqs.~(\ref{Pt}) and 
~(\ref{rhoPr}), 
\begin{eqnarray*} 
\rho &\approx& v^{-p-1} e^{-\alpha_0}\bigl(A_0(u) +qB_0(u) \bigr) \\ 
P_r &\approx& v^{-p-1} e^{-\alpha_0}\bigl(-A_0(u) +qB_0(u) \bigr) \\ 
P_{\perp} &\approx& -v^{-p-1} e^{-\alpha_0} \frac{q}{2} \beta'_0. 
\end{eqnarray*} 
If we require space-time to be radiation dominated in the neighborhood of 
$v=0$, then 
\[-\frac{q}{2} \beta'_0(u)= -A_0(u)+qB_0(u) = \frac{1}{3}\bigl(A_0(u)+qB_0(u) 
\bigr)\] 
which gives the readily satisfied conditions 
\begin{equation} \label{nullrad} \beta'_0(u)<0, \qquad A_0(u) = -\frac{q}{2} 
\beta'_0(u), \qquad B_0(u) = -\beta'_0(u). 
\end{equation} 
 
The conclusion is that for a spherically symmetrical solution, a power-law 
type singularity surface can occur for an isotropic radiative equation of 
state $P=\frac{1}{3}\rho$, which is either timelike or null, provided there 
is the simple relation $q=p+2$, with $q=4$ and $p=2$, and $q=p+1$ with 
$p>-1$ respectively in the leading exponents of $U$ and $V$ in 
Eq.~(\ref{eUV}). \\ 
 
An interesting property of these singularities is linked to their 
area whose magnitude is $4 \pi$ times the coefficient 
of $d\Omega^2$ in the expression for the metric. If the singularity has zero 
area, it can be considered as a central focus. If its area is 
finite, the singularity is usually regarded as being a shell-cross. 
In Eq.~(\ref{mdn}), the coefficient of $d\Omega^2$ is 
$e^V$, which, from the definition retained in Eq.~(\ref{eUV}), is 
equal to $x^q e^\beta$. Therefore, every singularity $x=0$ such that 
$q>0$ is a central focus. This is the case for both the timelike 
and null singularity solutions above identified. 
 
\section{Cosmological applications} 
\label{cissdm} 
 
We now turn our attention to the cosmological consequences we can derive 
from the above stated results. If we consider the physically consistent 
picture of a universe which is first radiation dominated and after a period of 
cooling, becomes dust dominated, we are now provided with two different ways of 
giving a final solution to the horizon problem, using the scheme of 
Fig.~\ref{f:2}. The first is to assume, as in \cite{MNC00,CS98,SC99}, 
that a consistent approximation of the dust dominated region of the 
universe can be a model pertaining to the DBB class. We shall discuss 
below some salient features of this class of models. The second is to 
take advantage of the new results to suggest that in a radiation 
dominated primordial universe timelike or null singularities can occur 
therefore getting rid of any horizon problem, whatever the properties of 
the dust dominated era to come. \\ 
 
One feature of the DBB model worth to be taken into account is the 
nature of the constant energy density surfaces. 
We have seen indeed, in Sec.~\ref{sol}, that causality is restored between 
every pair of points on the last-scattering surface, provided the backward 
light-cone issued from these points reconnect at the ``center'' of the model 
before reaching the singularity. This can be achieved by the virtue of a 
non-spacelike constant energy density surface interposed between the 
last-scattering surface and the singularity. In a pure dust DBB model, the 
shell-crossing singularity, which can be viewed as a surface of infinite 
``constant'' energy density, is timelike. We show, in the followings, that 
this timelike property is shared, in this model, by a set of constant 
high-energy density surfaces, but that less energetic such surfaces are 
spacelike. We are therefore induced to look for peculiar subclasses of DBB 
models for which the non-spacelike nature would be shared by a sufficently 
broad set of constant high-energy density surfaces such as to include 
surfaces with energy densities smaller than the limit where the dust and 
radiative energy densities are of the same order of magnitude. Such models 
would be free of any horizon problem, as one can convince oneself by 
replacing, in Fig.~\ref{f:2}, the non-spacelike singularity by a 
non-spacelike (timelike) constant density surface.\\ 
 
In the dust dominated region of a DBB model, corresponding to a zero 
energy (spatially flat) Lema\^itre-Tolman-Bondi solution \cite{GL33,RT34,HB47}, 
the line element in comoving coordinates ($r,\theta,\varphi$) and proper 
time $t$, is: 
\begin{equation} 
d
s^2 = -c^2 dt^2 + R'{}^2(r,t)dr^2 + R^2 (r,t)(d\theta^2 + \sin^2 \theta 
d\varphi^2). \label{eq:ltb} 
\end{equation} 
 
With the radial coordinate $r$ defined as in \cite{CS98}, we obtain an 
expression for the metric component $R$, 
\begin{equation} 
R(r,t) = \left({9GM_0\over2}\right)^{1/3}r[t-t_0(r)]^{2/3},  \label{eq:gr} 
\end{equation} 
and for the energy density 
\begin{equation} 
\rho(r,t) = {1\over2\pi G[3t-3t_0(r)-2rt'_0(r)][t-t_0(r)]},  \label{eq:rho} 
\end{equation} 
where $t_0(r)$ is an arbitrary function of $r$, such that $t=t_0(r)$ is 
the focusing ``Big Bang'' singularity surface for which $R(r,t)=0$. \\ 
 
We see from Eq.~(\ref{eq:rho}) that the equation for the surfaces with 
constant energy density can be written 
\begin{equation} 
D(r,t) = [3t-3t_0(r)-2rt'_0(r)][t-t_0(r)]=\text{const}.\label{eq:const} 
\end{equation} 
 
The normal, $n_{\beta}$, to this surface is 
\begin{equation} 
n_{\beta} \propto (\dot{D},D',0,0), \label{eq:nb} 
\end{equation} 
where a dot denotes the derivative with respect to $t$, and a prime, the 
derivative with respect to $r$. \\ 
 
From Eq.~(\ref{eq:const}), we get 
\begin{equation} 
\dot{D} = 2(3t-3t_0-rt'_0),\label{eq:dd} 
\end{equation} 
\begin{equation} 
D' = 2[r{t'_0}^2-(4t'_0+rt''_0)(t-t_0)],\label{eq:dp} 
\end{equation} 
and substitution into Eq.~(\ref{eq:nb}), after simplifying by the constant 
factor $2$, results in 
\begin{equation} 
n_{\beta} \propto (3t-3t_0-rt'_0,r{t'_0}^2-(4t'_0+rt''_0)(t-t_0),0,0). 
\label{eq:nbd} 
\end{equation} 
 
Using the metric tensor components as they appear in Eq.~(\ref{eq:ltb}), we 
can write 
\begin{equation} 
n_{\beta} n^{\beta} = -c^2(3t-3t_0-rt'_0)^2 + R'{}^2[r{t'_0}^2- 
(4t'_0+rt''_0)(t-t_0)]^2.\label{eq:nbnb} 
\end{equation} 
It can now be verified that on the shell-crossing surface $R'=0$, 
corresponding 
to $3t-3t_0-2rt'_0=0$ \cite{CS98}, the vector magnitude $n_{\beta} n^{\beta}$ 
is negative for every value of $r$ and $t$, confirming the timelike nature of 
this surface (see the Appendix). \\ 
 
We now consider a constant energy density surface located in the $R'>0$ 
region, with $\rho$ sufficiently large to be allowed to write: 
\begin{equation} 
3t-3t_0-2rt'_0 = {1\over{2\pi G \rho (t-t_0)}}=\epsilon (r,t), 
\label{eq:eps} 
\end{equation} 
with $0<\epsilon(r,t) \ll rt'_0(r)$ for every $r$ and $t$. In this 
case, 
\begin{equation} 
n_{\beta} n^{\beta} = -c^2(\epsilon + rt'_0)^2 + \left({GM_0\over 2}\right) 
^{2/3}{\epsilon^2\over(\epsilon + 2rt'_0)^{2\over 3}}\left[ 
r{t'_0}^2 - (4t'_0+rt''_0)({\epsilon+2rt'_0\over 3})\right]^2,  \label{eq:tpsc} 
\end{equation} 
where the right hand side is dominated by the negative term of zero order in 
$\epsilon$, namely $-c^2r^2{t'_0}^2$.
The corresponding surface of constant $\rho$ is therefore timelike.\\ 
 
On the other hand, a constant low energy density surface, satisfying 
\begin{equation} 
3t-3t_0-2rt'_0 = {1\over{2\pi G \rho (t-t_0)}}={1\over \epsilon(r,t)}, 
\label{eq:useps} 
\end{equation} 
where $0<\epsilon(r,t) rt'_0(r)\ll 1$ for every $r$ and $t$, gives, after 
an expansion in powers of $\epsilon^{-1}$, 
\begin{equation} 
n_{\beta}n^{\beta} = -{c^2\over \epsilon^2}(1+\epsilon rt'_0)^2 + 
\left({GM_0\over 2}\right)^{2/3}{(4t'_0+rt''_0)^2\over{9\epsilon^{10\over 
3}}}[1-{\cal O}(\epsilon rt'_0)][1-{\cal O}(\epsilon rt'_0)]^2. \label{eq:tlsc} 
\end{equation} 
The positive term of $10\over 3$th order in $\epsilon^{-1}$ dominates 
in this equation, and the corresponding surface of constant $\rho$ is therefore 
spacelike. \\ 
 
In \cite{MNC00,CS98} the physical assumption is made that the surface of 
last-scattering is a spacelike constant temperature (i.e., constant low energy 
density) surface, located in the dust dominated era. 
When traveling backward on an incoming light-cone emitted from any point on 
this surface, we therefore cross the spacelike $\rho =$ const.\ surfaces 
exhibiting growing energy densities until we reach either the region where 
these surfaces become timelike or the radiation dominated era. \\ 
 
If the first timelike $\rho =$ const.\ surface is still located in the dust 
dominated region, the lightcone is bound to reconnect at the center before 
reaching this surface and the horizon problem naturally disappears for any 
observer looking at any point on the last-scattering surface. If the 
radiation dominated domain is reached first, we can contemplate two 
possibilities. \\ 

Firstly, we may be brought back to an inflation-like configuration 
(see Fig.~\ref{f:1}), where the horizon problem can be (temporarily) solved 
for a given observer, provided the backward incoming light-cone issued from 
the ``point'' (i.e., two-surface) she sees on the last-scattering 
spends enough space-time in the dust dominated era. 
This implies some fine tuning of the parameters of the model, i.e. the observer 
location $r_0$ and the expression for the $t_0(r)$ function, of which the 
essential variable is the slope. Although the allowed values for these 
parameters can be chosen from related infinite sets, such a solution 
provides less intellectual satisfaction than our alternative proposal. 
We need a means to discriminate between DBB models which do or do not exhibit
timelike $\rho_d$ surfaces such that $\rho _d<\rho_{eq}$, where $\rho_{eq}$ 
is the value of the energy density for which the dust $\rho _d$ and radiation 
components are equivalent. While this cannot be done analytically, 
the problem can easily be solved numerically for any given profile 
of the ``Bang'' function $t_0(r)$ (see, e.g., \cite{CS98} where two examples 
have been numerically solved). \\ 

Alternatively, the radiation dominated region may be smoothly 
connected to the dust dominated region by using a model among the solutions of 
Einstein's equations exhibiting a singularity of the timelike or null type 
identified in Secs.~\ref{ts} and ~\ref{ns}. It remains a problem, however, 
that if such solutions exist, we do not have sufficient knowledge of 
their other properties to provide a matching of these solutions with the 
DBB ones. \\ 
 
Notwithstanding these difficulties, it seems to us that the simplest way of 
resolving the horizon problem is to take 
advantage of the new results stated in the present article and only consider 
the primordial region of the universe, i.e., the neighborhood of the 
singularity which we can physically assimilate to an era of energy density 
approaching the Planck scale. If this region can be represented by one of 
the radiative models identified in Secs.~\ref{ts} and ~\ref{ns} as 
exhibiting a timelike or null singularity, then the problem is 
definitely solved. 
 
\section{Application to the SCCH} 
\label{scch} 
 
The Strong Cosmic Censorship Hypothesis (SCCH) was proposed in 1979 
by Penrose \cite{RP79} after the debate which had followed his first 
proposal, in 1969, of the Cosmic Censorship Hypothesis 
\cite{RP69}. The SCCH runs as follows: \\ 
{\it No physically realistic collapse leads to a locally naked (i.e., 
timelike) singularity}. \\ 
These singularities are visible from regular points of space-time, 
but possibly not at infinity. However, from the point of view of 
infalling particles, such singularities must be as worrying as 
those visible at infinity, since they are likely to upset the 
physical conditions in their space-time neighborhood. (see, e.g.\ SI 
for a further discussion of this issue). \\ 
 
In Sec.~\ref{ts}, we have identified a class of power-law type 
focusing timelike singularities, with spherical symmetry, 
exhibiting in their vicinity the stress-energy tensor of a radiative 
perfect fluid. Reversing the sign of time, we obtain a 
corresponding class of solutions to Einstein's equations which 
represent, to a good approximation, a 
spherical cloud of collapsing gas near its focusing point. It 
therefore constitutes an interesting, physically 
consistent counter-example to the SCCH. 
 
\section{Conclusion} 
\label{concl} 
 
In this paper, we have extended the study, initiated in SI, of spherically 
symmetric metrics satisfying the dominant energy condition and of which the 
singularities are of power-law type. We have identified two classes 
of peculiar interest: 
 
\begin{enumerate} 
\item 
 
A {\it timelike} class exhibiting in the neighborhood of its {\it 
focusing} singularity the stress-energy tensor of a perfect fluid, with 
the equation of state of {\it radiation}: $p={1\over 3} \rho$. 
 
\item 
 
A set of {\it null} classes verifying identical properties. 
 
\end{enumerate} 
 
We have considered two important applications of these results: 
 
\begin{enumerate} 
 
\item 
{\em In cosmology, the possibility of solving the horizon problem.}  
We have reviewed in Sec.~\ref{sol} how a timelike or null singularity 
is a sufficient condition for a cosmological model to become rid of 
this cumbersome problem. Therefore, if we consider the physically 
consistent picture of a universe which is first radiation dominated and, after 
a period of cooling, becomes dust dominated, we can take advantage of our new 
results to state that Einstein's equations permit the existence of solutions 
exhibiting non-spacelike singularities having  physical conditions in their 
neighborhood consistent with the primordial region of 
the universe. These can be assimilated to a region of energy density 
approaching the Planck scale (beyond which General Relativity is 
generally believed to break down). Choosing to describe this region with one 
of the radiative models corresponding to the timelike or null 
singularities identified in Secs.~\ref{ts} and ~\ref{ns} allows us to solve 
{\it permanently} the horizon problem, as has been stressed in 
Sec.~\ref{sol}. \\ 

Together with the DBB model \cite{CS98}, first proposed to solve 
the horizon problem in a geometrical way, these results provide us with 
new candidates to achieve this without need to resort to an 
inflationary scenario. It is worth noting that, contrary to 
the DBB solutions which exhibit a shell-crossing singularity, those 
proposed here arise from a {\it focus}. 
 
\item 
{\em In gravitational collapse, a counterexample to the SCCH.}
If we limit ourselves to the consideration of the focusing 
{\it timelike} singularities identified in Sec.~\ref{ts}, the 
corresponding class of solutions to Einstein's 
equations represents, to a good approximation, a 
spherical cloud of collapsing gas near its focusing point, contradicting the
commonly believed Strong Cosmic Censorship Hypothesis.
 
\end{enumerate} 
 
Some further cosmological conclusions can incidentally be derived 
from the results first obtained in SI. We recall that, in that
work, it was shown that in the neighborhood of a timelike 
singularity of power-law type which is not a shell-cross, the dominant energy
condition can be satisfied if there is an
asymptotically extreme energy-stress tensor ($|P_r|\sim \rho$ 
or $|P_{\perp}|\sim \rho$) or one of the pressures is negative.
As a negative pressure is characteristic of a 
cosmological constant dominated universe, such a model 
is not likely to exhibit any horizon problem. \\ 
 
We finish by noting that the timelike nature of any spherically 
symmetric shell-crossing singularity, stated in the Appendix, 
would allow to extend from the DBB cosmological models to {\it non-flat} 
predominantly-dust cosmological models the property of solving the 
horizon problem. We leave the discussion of such models to future 
works, stressing once more the nice geometrical properties 
possessed by the geodesics in some peculiar classes of 
{\it inhomogeneous} models of universe.

\section{Appendix} 
 
In this Appendix, we give a derivation of the timelike character of the 
shell-crossing singularity in a spherically symmetric model, generalizing 
a line of reasoning first proposed by Hellaby and Lake in \cite{HL85}. \\ 
 
The general spherically symmetric line element, can be written: 
\begin{equation} 
ds^2 = -B^2(r,t) dt^2 + A^2(r,t)dr^2 + R^2(r,t)(d\theta^2 + 
\sin^2 \theta d \varphi^2) .  \label{eq:1} 
\end{equation} 
 
A typical shell-crossing surface $t=b(r)$ is such that 
\begin{equation} 
A =[t-b(r)]^a f(r,t)=0\qquad B\neq 0 \qquad R\neq 0. \label{eq:2} 
\end{equation} 
 
The normal, $n_{\alpha}$, to the surface $A=$const. (here $A=0$), is 
\begin{equation} 
n_{\alpha} \propto (\dot{A},A',0,0). \label{eq:3} 
\end{equation} 
 
With the metric of Eq.(\ref{eq:1}), the squared norm of this normal vector is 
\begin{equation} 
n_{\alpha} n^{\alpha} =-{\dot{A}}^2 B^2 + {A'}^2 A^2. 
\label{eq:7} 
\end{equation} 
 
According to Eq.(\ref{eq:2}), $A^2=0$, and the above 
expression for $n_{\alpha}n^{\alpha}$ is always negative, implying, with 
our choice of the metric signature, a timelike shell-crossing surface.

\end{document}